\documentclass{PoS}

\usepackage{booktabs}
\usepackage{float}

\newcommand{\reg}{\textsuperscript{\tiny{\textregistered}}}
\newcommand{\tm}{\textsuperscript{\tiny{\texttrademark}}}

\title{ASTRI SST-2M prototype and mini-array simulation chain, data reduction software, and archive in the framework of the Cherenkov Telescope Array}

\ShortTitle{ASTRI simulation chain, data reduction software, and archive}

\author{
S. Lombardi$^{1,2}$, C. Bigongiari$^{1,2}$, S. Gallozzi$^{1}$, L.A. Antonelli$^{1,2}$, D. Bastieri$^{3}$, I. Donnarumma$^{4}$, F. Lucarelli$^{1,2}$, M. Mastropietro$^{1,2}$, P. Munar$^{5}$, M. Perri$^{1,2}$, A. Stamerra$^{2,6,7}$, F. Visconti$^{1,2}$, M.C. Maccarone$^{8}$ for the CTA ASTRI Project$^{9,10}$\\
\footnotesize{
(1) INAF--OAR, Rome (Italy); (2) ASI Science Data Center, Rome (Italy); (3) University of Padua (Italy); (4) INAF--IAPS, Rome (Italy), now at ASI, Rome (Italy); (5) INAF--IAPS, Rome (Italy); (6) INAF--OATO, Turin (Italy); (7) Scuola Normale Superiore, Pisa (Italy); (8) INAF--IASF, Palermo (Italy);\\ 
(9) http://www.brera.inaf.it/astri/; (10) http://www.cta-observatory.org\\
E-mail: \email{saverio.lombardi@oa-roma.inaf.it}; \email{ciro.bigongiari@oa-roma.inaf.it}; \email{stefano.gallozzi@oa-roma.inaf.it}
}
}

\abstract{
The Cherenkov Telescope Array (CTA) is a worldwide project aimed at building the next-generation ground-based gamma-ray observatory. CTA will be composed of two arrays of telescopes of different sizes, one each in the Northern and Southern Hemispheres, to achieve a full sky-coverage and a ten-fold improvement in sensitivity over an unprecedented energy range extending from 20~GeV to 300~TeV. Within the CTA project, the Italian National Institute for Astrophysics (INAF) is developing an end-to-end prototype of the CTA Small-Size Telescopes with a dual-mirror (SST-2M) Schwarzschild-Couder configuration. The prototype, named ASTRI SST-2M, is located at the INAF ``M.C. Fracastoro'' observing station in Serra La Nave (Mt. Etna, Sicily) and is currently in the scientific and performance validation phase. A mini-array of (at least) nine ASTRI telescopes has been then proposed to be deployed at the Southern CTA site, by means of a collaborative effort carried out by institutes from Italy, Brazil, and South-Africa.
The CTA/ASTRI team is developing an end-to-end software package for the reduction of the raw data acquired with both ASTRI SST-2M prototype and mini-array, with the aim of actively contributing to the global ongoing activities for the official data handling system of the CTA observatory. The group is also undertaking a massive Monte Carlo simulation data production using the detector Monte Carlo software adopted by the CTA consortium. Simulated data are being used to validate the simulation chain and evaluate the ASTRI SST-2M prototype and mini-array performance. Both activities are also carried out in the framework of the European H2020-ASTERICS (Astronomy ESFRI and Research Infrastructure Cluster) project. A data archiving system, for both ASTRI SST-2M prototype and mini-array, has been also developed by the CTA/ASTRI team, as a testbed for the scientific archive of CTA. The archive system provides the data access at different user levels and for different use cases, each one with a customized data organization. A dedicated framework to access, browse and download data produced by the ASTRI telescopes has been developed within a scientific gateway utility. In this contribution, we present the main components of the ASTRI data handling systems and report the status of their development.
}

\FullConference{35th International Cosmic Ray Conference --- ICRC2017\\
                10--20 July, 2017\\
                Bexco, Busan, Korea}

\begin{document}

\section{Introduction}
In the last decade, the firm detection of more than 160 galactic and extragalactic sources by the current generation of ground-based imaging atmospheric Cherenkov telescopes (IACTs)~\cite{hinton09}~--~H.E.S.S., MAGIC, and VERITAS~--~has demonstrated the huge physics potential of very high-energy (VHE, E~$\gtrsim$~50~GeV) gamma-ray astronomy~\cite{aharonian13}. The Cherenkov Telescope Array (CTA)~\cite{actis11} represents the new generation of IACT facilities and will dramatically boost the current IACT performance, widening the VHE science~\cite{acharya13}. In order to cover the wide energy range from a few tens of GeV up to a few hundreds of TeV, three different classes of Cherenkov telescopes having different sizes~--~large (diameter D$\sim$23~m); medium (D$\sim$12~m and D$\sim$9.5~m); small (D$\sim$4~m)~--~will be used.\\
ASTRI ({\it Astrofisica con Specchi a Tecnologia Replicante Italiana})~\cite{pareschi16} is a sub-project within CTA led by the Italian National Institute for Astrophysics (INAF). The primary goal of the ASTRI project is the realization of an end-to-end prototype of the CTA Small-Size Telescopes with a dual-mirror (SST-2M) Schwarzschild-Couder configuration. The prototype, named ASTRI SST-2M and located in Serra La Nave (Mt. Etna, Sicily), is currently undergoing the scientific verification phase\footnote{The first Cherenkov events have been successfully acquired in May 2017. Data taking on regular basis is foreseen by summer 2017.}. When this phase is complete, the project aims at deploying, for the CTA Southern site pre-production phase, a set of (at least) nine ASTRI telescopes, hereafter named the ASTRI mini-array. A description of the ASTRI telescopes and overall project is given elsewhere in these proceedings~\cite{maccarone17}. The ASTRI mini-array will be able to perform early CTA science~\cite{vercellone15} and test the ASTRI data handling system proposed to be part of the one eventually adopted by the CTA observatory. As an end-to-end project, ASTRI includes (besides all hardware and control software systems) the full data processing and archiving chain, from raw data up to final scientific products. In the following sections, we present the main components of the ASTRI data handling systems~--~namely, the data reduction software, the Monte Carlo (MC) simulation chain, and the scientific archive system~--~and report the status of their development, along with the main activities on-going for their validation.
%

\section{ASTRI data reconstruction and scientific analysis software}
The ASTRI data reconstruction and scientific analysis software (henceforth {\it A-SciSoft}) is the official package of the ASTRI project being developed for the ASTRI SST-2M prototype and mini-array data processing, in compliance with the general CTA requirements~\cite{lamanna15}. The software can handle real and MC data and is intended to provide all necessary algorithms and analysis tools for characterizing the scientific performance of the ASTRI SST-2M prototype and afterwards to carry out the foreseen mini-array scientific program~\cite{vercellone15}.\\ 
The main purpose of {\it A-SciSoft} is to provide all necessary software tools to reconstruct the physical characteristics of astrophysical gamma rays (and background cosmic rays) from the raw data generated by the ASTRI telescopes. The software is composed by a set of independent modules organized in efficient pipelines that implement the algorithms to perform the complete data reduction, from raw data to the final scientific products. The FITS data format~\cite{pence10} has been adopted for all ASTRI data levels (DL) and types\footnote{It is already foreseen that, in the case of the mini-array, the DL0 data will be likely handled in their original raw (rather than FITS) format.}, from DL0 up to DL4.
The code has been conceived to be easily ported to parallel computing architectures (see section~\ref{lp}), such as multi-core CPUs and graphic accelerators (GPUs), and new hardware architectures based on low-power consumption processors (e.g. ARM). C++ and Python have been chosen as the programming languages for the main components of the software suite. NVIDIA\reg{} CUDA\reg{} has been used to port the most computationally demanding algorithms to GPUs, while Python was used to wrap the executable modules in efficient pipelines. The final scientific products are then obtained by means of either science tools currently being used in the CTA Consortium (e.g. {\it ctools}~\cite{knodlseder16b}) or specifically developed ones. To deploy {\it A-SciSoft}, a conda package has been created making it very easy to install along with its dependencies and run on Linux and MacOSX machines.\\
A comprehensive description of high-level requirements, data model, data flow, functional design, and framework of the {\it A-SciSoft} software package can be found in~\cite{lombardi16}.
%

\section{ASTRI data reduction on low-power and parallel architectures} \label{lp}
We implemented the complete analysis chain required by a single telescope on a workstation equipped with a dual Intel\reg{} Xeon\reg{} E5-2650 CPU (8 physical cores and 95 W of Thermal Design Power each\footnote{See technical specifications here: http://ark.intel.com/products/64590/Intel-Xeon-Processor-E5-2650-20M-Cache-2\_00-GHz-8\_00-GTs-Intel-QPI}) and on a NVIDIA\reg{} Jetson\tm{} TX1 development board (quad-core ARM\reg{} Cortex\reg{}-A57 CPU). In this section we focus on the Jetson, which can complete the entire pipeline in $\sim$25~s on a single typical data-taking run constituted by $\sim$5.5$\times$10$^4$ events ($\sim$500~MB), while staying within a very small power budget \cite{mastropietro16}.\\
Considering the constraining requirements of the on-line analysis defined by CTA, we explored the possibility of creating a custom implementation on low-power (potentially embedded) hardware that could allow each telescope to process its own data before sending them to a central acquisition system. We re-factored the calibration and cleaning analysis steps to minimize latencies, and selected the Jetson TX1~--~offering both low power consumption and a CUDA-capable GPU~--~as the target platform for our experiment. We successfully tested the Jetson TX1 as an embedded module capable of carrying out the whole single telescope data processing pipeline, including signal calibration, image cleaning, image parameterization, single-telescope reconstruction, and event list production~\cite{lombardi16}. Leveraging its efficient system on chip (SoC), we found that it can process more than 2k events/s, twice as much as the maximum data acquisition (DAQ) rate, with a power consumption as low as 10~W.
We ran the entire pipeline on a single run of the data challenge ON sample, as described in section \ref{dc}. Execution time is higher on the low-power board by a factor 1.5. This can be improved by performing reconstruction operations (mainly Hillas parameters computation) on the GPU as well. However, the energy-to-solution is lower by more than an order of magnitude on the Jetson ($253$ J vs $2964$ J). This estimate is an upper bound, since we took the nominal power consumption of both the workstation and Jetson and multiplied it by the execution time.
%

\section{ASTRI simulation chain} \label{mc}
Simulations are an essential component of any IACT project. They are used for the design phase, for the commissioning, and during their operational period. The design optimization of any IACT heavily relies on simulated data. During the commissioning phase an iterative comparison between acquired and simulated data leads at the same time to the validation of the simulation chain and to the optimization of the operational parameters which affect the telescope performance. Simulations are mandatory also during the operational phase to train the background rejection strategies and to estimate the instrument response functions. ASTRI is no exception to this rule and large samples of simulated data have been already produced for the above mentioned purposes and to test the data reconstruction and scientific analysis software.\\
For the ASTRI simulation productions, the same simulation chain commonly used by the CTA consortium are adopted for easiness of comparison and consistency. Atmospheric showers are simulated using the {\it CORSIKA} code~\cite{corsika} (version 6.99) while telescope response are simulated with the {\it sim\_telarray} package~\cite{simtel}, which propagates photons hitting the primary mirror through the telescope optical system to the camera and simulates the photon detection, the electronic response, and the trigger logic. The proper simulation of the electronic chain of the ASTRI camera has been verified by comparison with a custom-written simulation code. Both gamma-ray and background events have been simulated to test the {\it A-SciSoft} software package and to estimate the performance of the ASTRI SST-2M prototype and the ASTRI mini-array~\cite{lombardi16}. Due to the much larger background flux with respect to that of any known gamma-ray source and to the very efficient background rejection achieved by the IACTs, very sizable samples of background events are needed to derive performance and validation quantities\footnote{The simulation of atmospheric showers could be realized exploiting distributed computing and in particular the GRID technology and the DIRAC framework \cite{ctadirac}.}.\\
One particular sample of simulated data, the so-called ASTRI MC release 003 (REL003 henceforth), has been used to test the whole ASTRI reconstruction and analysis chain (see section~\ref{dc}). For this production, $\sim$2.5$\times$10$^{7}$ gamma-ray and $\sim$7$\times$10$^{8}$ proton primaries were simulated, with energies distributed according to a power law of spectral index -2.0 between 100~GeV and 330~TeV (for gammas) and 600~TeV (for protons). Gamma-ray primaries were simulated as coming from a point-like source, at $20^{\circ}$ zenith angle and $180^{\circ}$ azimuth angle, while proton incoming directions were randomized within a cone with $6^{\circ}$ radius centered on the same direction. To increase the available  number  of events, while introducing a negligible statistical bias, the impact points of primaries on the observational plane were randomized  several times (10 times for gammas and 20 times for protons) according to a uniform distribution within a circle with radius equal to 1200~m for gammas and 2000~m for protons~\cite{mcsimulations}. An array composed by four concentric square rings of ASTRI telescopes plus one more telescope at the center of the squares (33 telescopes overall)  was then simulated, with inter-telescope distance ranging from 200~m to 350~m. Such arrangement allows us to study the performance of different telescope array layouts. All the main hardware characteristics of the ASTRI telescopes (optical system, camera, electronics) were simulated according to the most updated measurements. In particular, the simulation of the ASTRI camera electronic chain has been performed to reproduce as closely as possible the behavior of the adopted CITIROC ASIC~\cite{sottile16}.
%

\section{ASTRI data challenge} \label{dc}
\begin{figure} [ht]
\begin{center}
\begin{tabular}{c}
  \includegraphics[height=5.6cm]{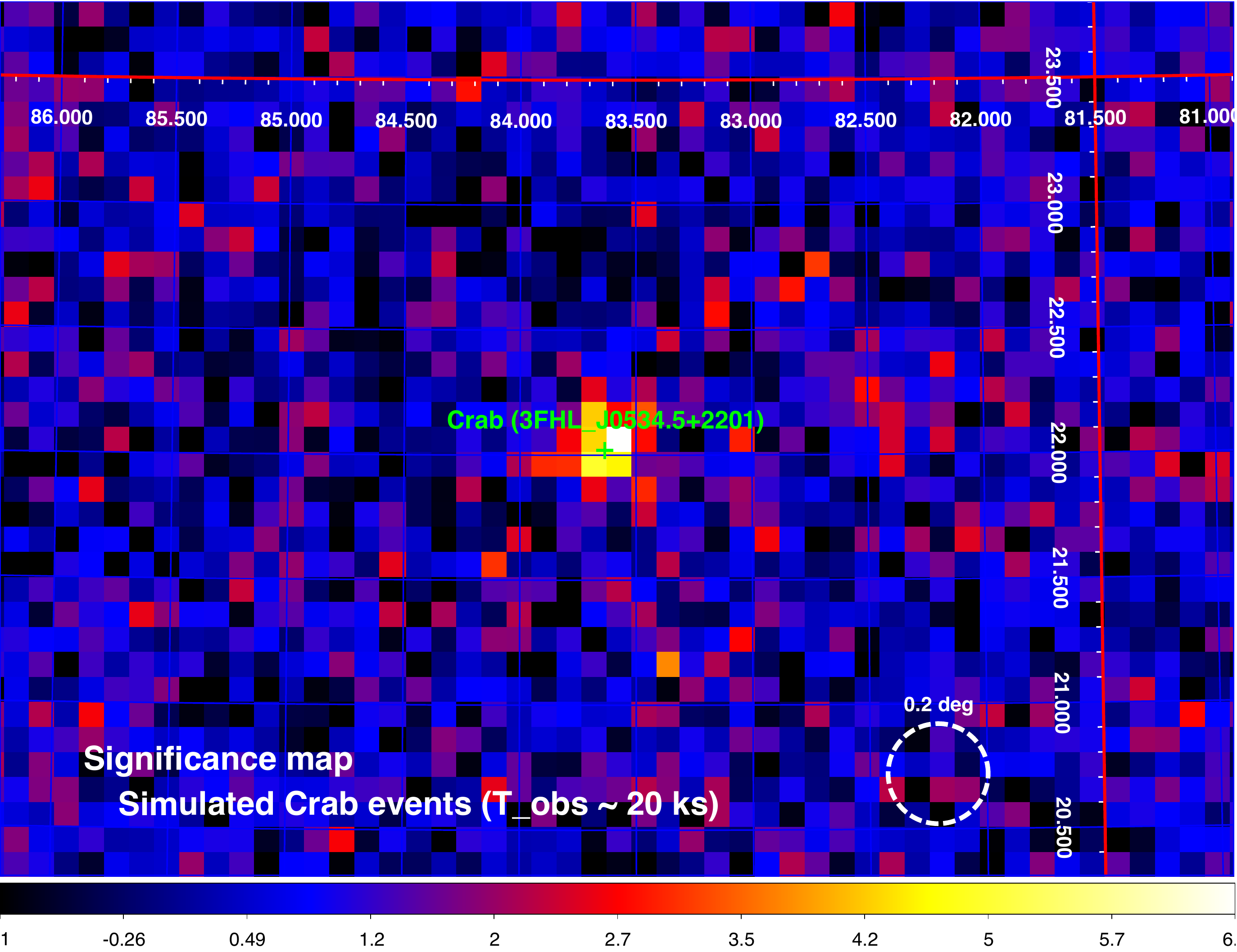}
  \includegraphics[height=6.1cm]{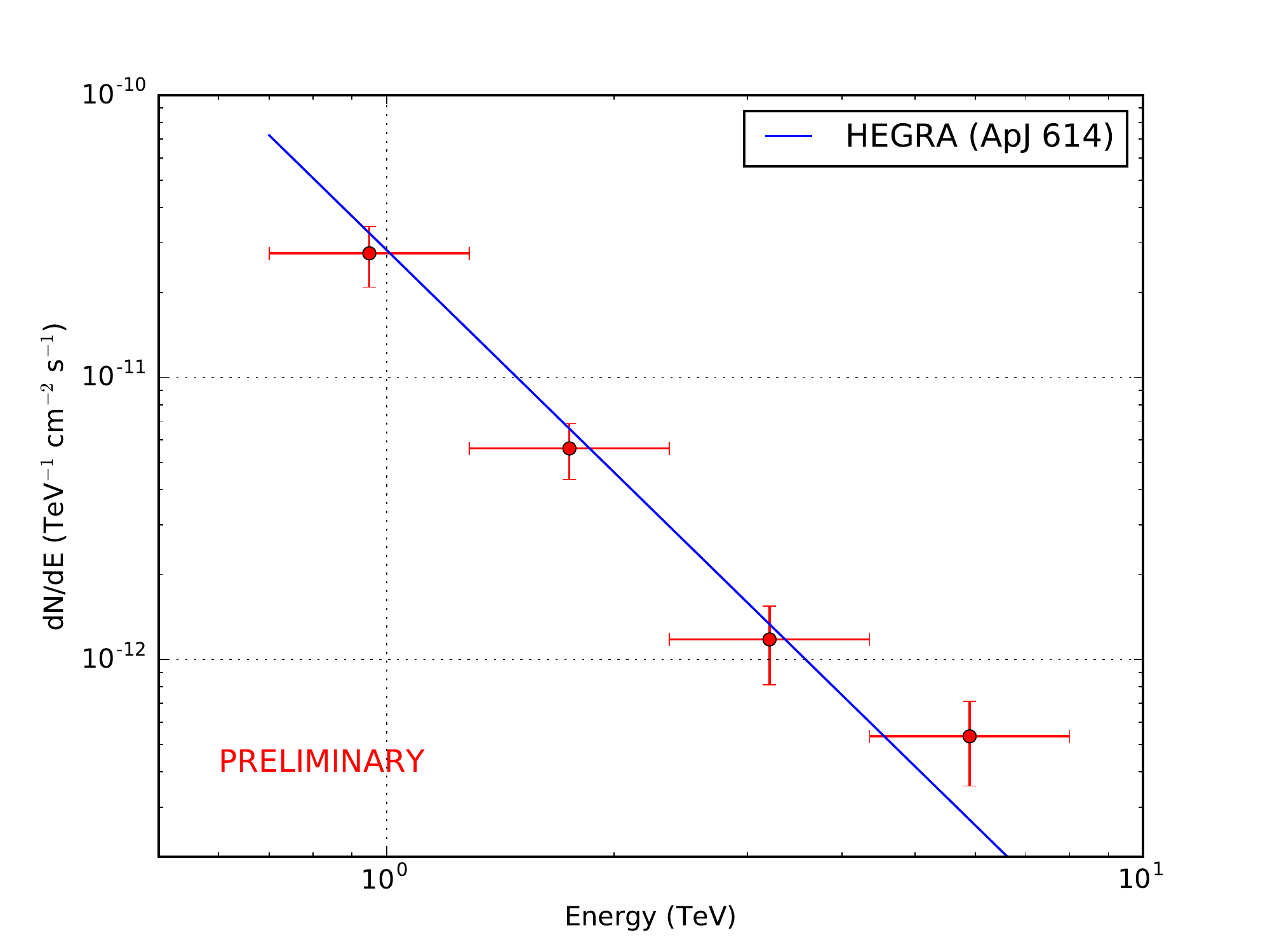}
\end{tabular}
\end{center}
\caption[]
{\label{figure1}
{\it Left}: Significance map of the ON data ($\sim$5.8 hours) sky region obtained with the {\it ctskymap ctools} task. The white dotted circle in the lower right indicates the point-spread function (68\% containment) of the analysis.
{\it Right}: Differential spectrum of the Crab Nebula between 0.7~TeV and 8~TeV obtained from the analysis of the ON data ($\sim$5.8 hours) obtained with the {\it csspec ctools} task. The blue line reprints the best-fit power-law parameterization of the Crab Nebula measured by the HEGRA Coll.~\cite{HEGRA}.
}
\end{figure}

In order to test the whole data reconstruction and scientific analysis chain in the case of single-telescope data reduction, a suitable DL0 data sample was created. The bulk of the scientific events of the data challenge sample was extracted from the REL003 ASTRI MC release, introduced in section~\ref{mc}. In order to maximize statistics, all MC proton events triggered by each of the 33 ASTRI simulated telescopes were put together at raw-data level (as if they were triggered by an ``average'' single ASTRI telescope). Then, the events were properly filtered so to follow the experimental energy slope of -2.70, as measured by the BESS Coll.~\cite{BESS}. After this procedure, the available statistics resulted in $\sim$4.2$\times$10$^6$ triggered events. Since the event rate was calculated to be $\sim$100~Hz, these amount of events corresponded to $\sim$11.6 hours of (single telescope) data taking. The overall data sample was then split in two subsamples, dubbed ``ON'' and ``OFF'', respectively. We combined the ON sample ($\sim$5.8 hours) with a number of randomly selected MC gamma-ray events calibrated to match the flux of the Crab Nebula as measured by the HEGRA Coll.~\cite{HEGRA}. The OFF sample, instead, was slightly reduced from $\sim$5.8 to $\sim$5.5 hours in order to keep a proper amount of independent proton events in their original MC format for the generation of gamma/hadron separation look-up-tables (LUTs). The instrument response functions (IRFs) were generated from MC gamma-ray events (for the effective areas, and angular/energy resolutions) and from the OFF sample (for the background rate). Finally, in order to associate a realistic time stamp and local arrival direction to each event, the ON and OFF data samples were assigned celestial coordinates assuming observations at the prototype site at Mt. Etna, Sicily, tracking the Crab Nebula (RA (J2000)~=~5h34m31.94s and Dec (J2000)~=~+22$^\circ$00'52.2'') and a close empty-sky region (RA (J2000)~=~5h50m36s and Dec (J2000)~=~+22$^\circ$33'51''). Both ON and OFF data were formatted at raw-data level in compliance with the ASTRI prototype (FITS) format. In particular, the data samples were divided in runs of $\sim$500~MB, each one containing $\sim$5.5$\times$10$^4$ events, as foreseen in real data taking condition. In the end, the ON and OFF data samples resulted in 39 ($\sim$5.8 hours) and 37 runs ($\sim$5.5 hours) of data taking, respectively. Along with the scientific ON and OFF raw data, all the other run-based DL0 inputs~\cite{lombardi16} were generated under realistic assumptions.\\
The ON and OFF data samples were successfully reduced from DL0 up to the generation of scientific products with the {\it A-SciSoft} software package. From DL0 to DL3, an efficient pipeline, wrapping the modules needed to perform the whole single telescope data reduction\footnote{More precisely, the {\it A-SciSoft} modules exploited for the reduction of the ON and OFF data samples were in sequence: {\tt astrical} (data calibration); {\tt astricleanpar} (image cleaning and parameterization; pointing and source position reconstruction); {\tt astrireco$^{ST}$} (full events' single-telescope-wise reconstruction); {\tt astrimer} and {\tt astrireco$^{A}$} (both acting as data format converters, because of the single-telescope data reduction); {\tt astriana} (for the reduced event-list and IRFs generation). A detailed description of the step-by-step ASTRI data reduction can be found in~\cite{lombardi16}.}, was used. Then, the DL3 data (reduced ON and OFF event lists and associated IRFs) were analyzed by means of the ASTRI science tools, in order to extract basic scientific products, like detection plots, sky-maps, and spectra. The {\it ctools} were successfully used to cross-checks the results. The left and right panels of Figure~\ref{figure1} show the significance sky-map (obtained by means of the {\it ctskymap ctools} task) and the Crab differential spectrum (using the {\it csspec ctools} task), respectively. Both results were achieved from the ON event list and associated IRFs, applying gamma/hadron cuts with 80\% efficiency for gamma rays. The results are basically in line with the performance expected for a single ASTRI telescope in the case of Crab Nebula observations and demonstrate the proper behavior of the whole data reduction chain. In particular, we found in a fiducial signal region of 0.2$^{\circ}$ around the source a significance of $\sim$7$\sigma$ (calculated according to the Eq.~17 of~\cite{LiMa83}) in $\sim$5.8 hours, which corresponds to a sensitivity of $\sim$25\% of the Crab Nebula flux in 50~hours of observations.
%

\section{ASTRI scientific archive}
The ASTRI SST-2M prototype scientific archive is identified with the off-site data archive, since the on-site archive is expected to be responsible only for buffering data for later off-site analysis and validation. The off-site data archiving and handling management system for both ASTRI SST-2M prototype and mini-array are developed within INAF institutes and take care of the full data chain produced by the different scientific devices at different analysis steps. The archive system provides the data access at different user-levels and for different use-cases, each one with a customized data organization. A dedicated framework to access, browse and download produced data has been developed within a scientific gateway utility. A customized proposal handling/scheduler system and user access utility for the ASTRI SST-2M prototype have been included too. Every data product is closely linked to the global ASTRI scientific data model in order to easily search related pieces of information on other archive elements and subsystems.\\
Different kinds of interfaces have been developed in order to simplify the interaction and integration of core software and developer's modules. The customized data organization and data model allow users to identify the best way to access data according to their user story and use case. A direct database connection with MySQL\reg{} and MongoDB\reg{} (and different API for CouchBaseDB) is foreseen. Every log entry related to data archiving, reduction, handling, and publication is saved into a MongoDB collection in order to allow simple and direct searches on the data.\\
The end-to-end experience in the ASTRI SST-2M prototype and the knowledge of suitable solutions for the management of databases' and Big Data issues related to CTA favor a distributed system as the practical archive organization. The mini-array project archive will be used as technological pre-production testbed to be adopted in the final CTA production. A first demonstrator was developed within the European H2020-Indigo Data Cloud project. It makes use of OneData package solutions for developing a distributed system deployed as a federation of different storage entities (i.e. three storage nodes). These storage entities are included in the data-grid topology infrastructure in order to inherit CTA simulation computing model and be ready to evolve in the ``storage as a service'' cloud paradigm. Scientific CTA end-users can adopt CTA authentication and authorization services granting permissions to carry out tasks in the federated archive, while simulation and pipeline users may connect to the archive system using standard grid authentication mechanisms. Such a structured testbed may become a {\it de facto} standard for multi-purpose adaptive storage applications ready for several heterogeneous scientific communities.\\
More details about the ASTRI scientific archive can be found in~\cite{carosi16}.
%

\section{Summary}
ASTRI is an end-to-end project for the development of telescopes of the CTA small-size class. In addition to the hardware and control software systems, the project includes a full data processing and archiving chain, from raw data up to final scientific products. In this contribution, we have presented some basic aspects of the ASTRI simulations, data analysis, and scientific archive.\\
Monte Carlo simulations are indispensable during all phases of the development and operation of an IACT, from its design  to  the  analysis of acquired data. Large samples of atmospheric showers and their detection by ASTRI telescopes have been simulated using the same  simulation  chain adopted by CTA.  The simulation of peculiar aspects of the ASTRI telescopes has been carefully verified. The large amount of CPU and disk space needed for such simulations has motivated the use of the EGEE/EGI Computing Grid. A  sample of about 25 millions gamma events and about 700 millions proton events has been simulated to test the ASTRI data reconstruction pipeline.\\
For the ASTRI data processing, a dedicated software package ({\it A-SciSoft}) is being developed. The software will perform the whole ASTRI SST-2M prototype data reduction chain (up to the generation of scientific products) and is intended to be used for the ASTRI mini-array data processing. One of the software's design and implementation goals was keeping energy consumption as low as possible; our tests conducted on realistic data for the entire pipeline confirm a successful implementation for both standard architectures and low-power boards such as the NVIDIA Jetson TX1. In order to test its core components for the single telescope data reduction, a first ASTRI data challenge has been produced and its main scientific results, in line with the expectations, have been shown.\\
The ASTRI prototype project has been used as testbed for many technological archive solutions involved in long term storage, data mining and user access to a large amount of astronomical data. A first demonstrator for a distributed archive concept model as required by the CTA consortium was developed and will be tested in the pre-production phase of the CTA observatory within the mini-array ASTRI telescopes.\\
The solutions adopted so far for the development of the ASTRI data handling are compliant with the general CTA requirements. Nevertheless, new possible solutions are under investigation. In this respect, the foreseen evolution of the ASTRI project from the ASTRI SST-2M prototype to the ASTRI mini-array will represent a unique opportunity to test these solutions on a configuration very similar to the one that will be eventually realized at the CTA observatory. This will give us the opportunity to actively contribute to the global ongoing activities aimed at developing the official data handling system of the CTA observatory.
%

\acknowledgments
\footnotesize{
This work was conducted in the context of the CTA ASTRI Project. This work is supported by the Italian Ministry of Education, University, and Research (MIUR) with funds specifically assigned to the Italian National Institute of Astrophysics (INAF) for the Cherenkov Telescope Array (CTA), and by the Italian Ministry of Economic Development (MISE) within the ``Astronomia Industriale'' program. We acknowledge support from  the Brazilian Funding Agency FAPESP (Grant 2013/10559-5) and from the South African Department of Science and Technology through Funding Agreement 0227/2014 for the South African Gamma-Ray Astronomy Programme. We gratefully acknowledge financial support from the agencies and organizations listed here: http://www.cta-observatory.org/consortium\_acknowledgments/. This work has been supported by H2020-ASTERICS, a project funded by the European Commission Framework Programme Horizon 2020 Research and Innovation action under grant agreement n. 653477.
}



\end{document}